## Table1 :

| | $\phi$(eV) | $R_s(\Omega)$ | $R_p(\Omega)$ |
|---|---|---|---|
| **Virgin state** | 0.52 | 200 | $\propto$ |
| **LRS$_1$** $V<V^*$ $V>V^*$ | 0.35 0.33 | 60 30 | 280 300 |
| **LRS$_f$** $V<V^*$ $V>V^*$ | 0.35 0.22 | 60 7 | 280 300 |

**Voltage bias induced modification of all oxide $Pr_{0.5}Ca_{0.5}MnO_3/SrTi_{0.95}Nb_{.05}O_3$   junctions**

**Barnali Ghosh[*], K. Das  and A.K. Raychaudhuri[ᴵ]**

*Department of Materials Science, S.N.Bose National Centre for Basic Sciences,*

*Block-JD, Sector-III, Salt Lake, Kolkata-700 098, INDIA*

In this paper we report what happens to a pristine oxide junction $Pr_{0.5}Ca_{0.5}MnO_3/SrTi_{0.95}Nb_{.05}O_3$ (PCMO/Nb:STO), when it is subjected to cycling of voltage bias of moderate value (±4V). It is found that the initial cycling leads to formation of a   permanent state of lower resistance where the lower resistance arises predominantly due to development of a shunt across the device film (PCMO). On successive voltage cycling with increasing magnitude, this state transforms into states of successive lower resistance that can be transformed back to the initial stable state on cycling to below a certain bias. A simple model based on p-n junction with shunt has been used to obtain information on the change of the junction on voltage cycling. It has been shown that the observation can be explained if the voltage cycling leads to lowering of barrier at the interface and also reduction in series resistance. It is suggested that this lowering can be related to the migration of oxygen ions and vacancies at the junction region. Cross-sectional imaging of the junction shows formation of permanent filamentary bridges across the thickness of the PCMO after the pristine p-n junction is first taken through a voltage cycle, which would explain appearance of a finite shunt across the p-n junction.

Email:

[*]barnali@bose.res.in (corresponding author)

[ᴵ]arup@bose.res.in

## I. INTRODUCTION:

In recent years devices based on oxides have attracted attention for applications in areas like non-linear conduction elements,[1] switching, memory-resistors,[2] and Field-Effect Transistors.[3] Among these devices multilayers of oxides need special mention because they can form p-n junctions, Schottky junctions, FET and such related junction based devices. For example, Perovskite manganites and doped titanates and their multilayers show unique room temperature electric pulse induced resistance switching effect in which the resistance of the compound can switch between two stable resistive states.[4,5] Often resistive switching effect has a memory which makes them attractive for non-volatile storage applications.[6] In addition to resistive switching strong non –linear conduction has been found in these manganites at low temperature that can be linked to phenomena like hot-electron effect[7] and even nanoscale localized regions of low resistance state in a matrix of high resistance state has been created.[8]

In recent years there are intense efforts to study, understand and apply the resistive switching properties of these materials to make useful devices.[6,9] The mechanism of switching is a topic of intense current research research. A number of mechanisms have been suggested for this. Various suggested mechanism of resistive switching, till date, include field-driven lattice distortions,[4] Schottky barriers with interfacial states, electrochemical migration at the metal/oxide interface,[10,11] and phase separation etc.[12,13]

In the context of perovskite oxides it has been known for some time that oxygen can migrate under electric field in $ABO_3$ class of oxides at temperatures close to or higher than 300K.[14,15]

It appears that at room temperatures, field induced motion of oxygen plays an important role in modifying local conductance in oxides and such mechanism has been used to explain switching in oxide devices (both binary as well perovskite oxides) including memory switching.[6] Recently there has been direct imaging of oxygen migration by high resolution Transmission Electron Microscopy (TEM) in binary oxides.[16]

This paper explores what happens to an all oxide p-n junction device, when it is subjected to voltage cycles of relatively low bias (<5V). The devices were made from epitaxial films of p-type perovskite oxide $Pr_{0.5}Ca_{0.5}MnO_3$ (PCMO, x=0.5), grown on lattice matched single crystalline substrates of n-type $SrTi_{0.95}Nb_{0.05}O_3$ (Nb:STO). We observe that repeated voltage cycles can lead to change in a junction in a systematic way leading to substantial change in junction current (for a given bias) as well as the junction rectifying behavior. This can be interpreted as reduction of the effective barriers at the junctions along with opening- up of shunt conducting paths. Interestingly, the modifications that happen can be a precursor for unipolar switching in such devices, as discussed below. Typically such unipolar switching are generally seen in binary oxides as opposed to bipolar switching seen in perovskite oxides like PCMO which show bipolar switching behavior.

## II. EXPERIMENTAL:

Films of $Pr_{0.5}Ca_{0.5}MnO_3$ (PCMO,x=0.5) of thickness ~500nm were deposited on single crystalline Nb:STO substrates by using Pulsed KrF Excimer Laser (COMPex 201) with wavelength 248nm. Laser energy density was $1.8J/cm^2$ and repetition rate was 5Hz. The distance between target and substrate was 5cm. Oxygen pressure within the chamber was 30 Pascal and during deposition the substrate temperature was kept at $750^oC$. Post deposition in-situ annealing was done under oxygen atmosphere at $500^oC$ for 1hr. The quality of the film was tested through X-ray diffraction (XRD) with a $\theta-2\theta$ scan mode. PCMO films grow epitaxially with a (001) normal orientation on the Nb:STO substrate as shown in Fig.1. The lattice constant of the PCMO film grown on Nb:STO is $3.78A^o$ which compares very well to the bulk value of PCMO $3.8A^o$.

The top electrode of Ag (lateral dimension ~200 μm) was grown by vacuum evaporation through a hard mask. The Scanning Electron Microscope (SEM) images of the top view (with Ag top electrode) as well as the cross section of the device are shown in Fig. 2. Image of the cross section of the device Ag/PCMO/Nb:STO was prepared by ion milling is shown in Fig 2(a). The top view of the device is shown in Fig 2(b).

The Current–voltage (I–V) characteristics were measured in current – perpendicular to plane (CPP geometry), using a source-meter (Keithley; model 2400) and with microprobes shown schematically in the inset of Fig.2 (a). Evaporated Ag has been used as the electrode on both the oxides.

The resistivities of the PCMO films were characterized using measurements of in-plane resistivity by 4-probe technique using evaporated Ag contacts. For this work the PCMO films were grown on insulating single crystalline substrate $SrTiO_3$ (STO). The film on STO was grown simultaneously with the film on Nb:STO in the same growth chamber following the same growth parameters and heat treatments.

### III. RESULTS:

The in-plane I-V curve taken on a typical PCMO film is shown in Fig. 3 .The PCMO films have resistivity ~ 0.6 Ohm-cm. at room temperature. The evaporated Ag electrodes form a non-rectifying ohmic contact with almost linear I-V curves till bias up to ±4 volts. Weak non-linearity that sets in the observed I-V curves at higher bias, are similar to that seen in even single crystals at or near room temperatures[1] (Note. Single crystals of PCMO show strong non-linear behavior below the charge ordering transition temperature of ~260K. The non-linear behavior at lower temperature has been attributed to electronic reasons like hot-electron effect).[7]

All measurements were performed at room temperature. As stated before, the bias to the oxides has been applied through evaporated Ag. (The evaporated Ag was found to give nearly ohmic contact with the materials). The positive bias is defined with the top electrode PCMO as positive.

A typical I-V curve for the virgin PCMO/Nb:STO devices (PCMO positive ), before start of the

voltage cycling, has been shown in Fig. 4. The bias has been limited to < 1.5V. The I-V curve

shows a p-n junction type diode behavior with a series resistance. The line through the data is the

fit to the model of a p-n junction with a series resistance, discussed in the next section. Such a

junction type behavior has been seen in manganite – Nb:STO junctions which are either Schottky

type or with p-n junction type with hole-doped semiconductors like PCMO or metallic electrodes

like $La_{0.7}Sr_{0.3}MnO_3$ (LSMO).[17]

Next we studied the effect of voltage cycling with the maximum voltage excursion ($V_{max}$) being enhanced gradually. The voltage bias was scanned in the cycle as $0 \rightarrow +Vmax \rightarrow 0 \rightarrow -Vmax \rightarrow 0$.

Data are shown in Figure 5. The first jump from the virgin diode like behavior occurs at a

forward bias of $V \approx 1.9$ V (see Fig 5(a)). The jump leads to an enhancement of the device current

by nearly 110%-120%. The enhancement of current, as we will see below, occurs due to an

opening up of a low resistance shunt across the junction as well as reduction the barrier at the

junction. We call this state as $LRS_1$. The device stays in $LRS_1$ on voltage cycling (as long as

$V_{max}$ is < 2.2V) including swing to reverse bias (see Fig.5a). However, the $LRS_1$ state in the

reverse swing does not show rectifying behavior any more and appears to behave as an ohmic junction. We discuss this later on.

The gradual enhancement of the $V_{max}$ leads to another jump of current at $V \approx 2.2V$. The jump

leads to enhancement of current by another 180%-190%. (See Fig. 5(b)). We denote this lower

resistance state as $LRS_2$. On reducing the bias, the $LRS_2$ shows return to the $LRS_1$ at a voltage

$V \approx 0.8V$. We denote this voltage as $V^*$. For negative bias and for $V < V^*$, the I-V curve follows

that of the $LRS_1$ state and when the bias is enhanced in the positive swing, at $V = V^*$ it changes to the I-V curve of the $LRS_2$ state. On cycling the bias with $V_{max} < 2.8V$, the I-V curve follows a very reproducible and stable behavior. Interestingly, on increasing the bias $V_{max}$ further another jump in the current occurs for V≈2.9V, where the current changes by another 240%-250%. We call this $LRS_3$ (See Fig. 5(b)). The bias cycling with gradually enhanced $V_{max}$ shows a stable enhancement of current from the virgin state by a factor of nearly 10 (measured at V=1.5V). Again on reducing the bias the $LRS_3$ state changes back to the $LRS_1$ state at V≈$V^*$ and in the reverse bias region the I-V curve follows the I-V curve of the $LRS_1$ state closely. On cycling the bias, the device changes from $LRS_1$ to $LRS_3$ (and vice versa) reproducibly at V≈$V^*$.

This behavior of changing to successively lower resistance stated as Vmax is gradually increased, continues and the device eventually settles to a final state when $V_{max}$~3 volts. I-V curve of the final low resistance state (denoted as $LRS_f$) is shown in Figure 6 along with that of the I-V curve of the $LRS_1$ state. It shows that the bias cycling lead to nearly 25-30 fold enhancement of the current and that on reducing the bias, at V= $V^*$, the $LRS_f$ turns back to the $LRS_1$ state that is first formed from the virgin state.

To summarize the main observations, we find that to start with at low bias the device behaves like a simple rectifying diode. However, the diode state is not a stable state against voltage cycling and "switches" to a lower resistance state first ($LSR_1$) for $V_{max}$≈1.9V. This state is a stable state at low bias (V<$V^*$) and at negative bias and behaves like an ohmic junction. The device never goes back to the virgin rectifying state from the $LSR_1$ state. Though the $LSR_1$ state is stable at low bias and negative bias, on successively increasing the positive bias ($V_{max}$) the junction    goes through a series of transition to progressively lower resistance states and eventually reaching a final state ($LSR_f$) which is a stable low resistance state for bias V> $V^*$. For V< $V^*$, this state "switches" back to the stable $LSR_1$ state.  This observation of switching back and forth between two stable states ($LSR_1$ and $LSR_f$) occurs under unipolar condition and is distinct from bipolar switching of resistance states seen in almost all perovskite oxides. (The final two states are shown in Fig. 6.) We find that this behavior is retained over a long time (~$10^5$ sec or more). Below we propose a model and show that this behavior arises due formation of a current shunt that develops when the virgin diode is switched permanently to the $LRS_1$ and the shunt acts in parallel to the diode junction PCMO/Nb:STO, whose barriers also get modified by the bias. We also provide microscopic evidence for such shunts. The change of the virgin rectifying diode to the $LRS_1$ is very similar to the "formation" step seen in most unipolar resistive switching devices made on binary oxides .[18,19]

## V. DISCUSSIONS:

## V.1 MODEL

The observed behavior of the PCMO/Nb:STO device on voltage cycling is explained with a simple model of a p-n junction diode with a series resistance ($R_s$) and a shunt resistor ($R_p$) as shown in inset of Figure 6. The voltage cycling was found to change the junction barrier of the diode as well as the $R_s$ and $R_p$. In the virgin state the current through the diode (of area A) at temperature T in series with the resistor $R_s$ is given by Eqn. 1.[20]

$$I = A.A^*T^2 \exp\left(\frac{-q\Phi}{kT}\right)\left\{\exp\left[\frac{q(V - IR_s)}{nkT}\right] - 1\right\} \qquad (1)$$

With current being given by thermoionic emission for the reverse saturation current through a barrier of height $\Phi$, $A^*$ =Richardson constant, and n is the ideality factor. If we consider that the base materials (PCMO film as well as the Nb:STO substrates) present the series resistance $R_s$ to the diode formed by the junction of Nb:STO with PCMO. Eqn. 1 is a lumped equation where the current through the diode can be composed of contributions of a number of diodes in parallel formed at different places located over the area of the film junction with the substrate. The fit to the virgin state diode is shown as lines in Fig 4. The data can be fitted with a barrier height $\Phi$=0.52eV, the series resistance $R_s \approx$200 ohms and n=1.7. For the virgin device no shunt is needed ($R_p \rightarrow \infty$) and the whole current in the device flows through the p-n junction (see Table I).

We can fit the I-V data taken for different values of $V_{max}$ to the above model. While the data of the virgin diodes state does not need any shunt resistance $R_p$, for all other junctions obtained after successive voltage cycling would need a finite shunt path that is lumped into the parallel resistance $R_p$. This can be composed of a number of parallel resistances that shunts across the diode. The fit to the data are shown as solid lines in Fig. 6 for the two stable states (LRS$_1$ and LRS$_f$). The parameters obtained from the fit are shown in Table I.

The transition from the virgin state to the LSR$_1$ state that occurs with bias V~1.9V, as can be seen from Table I, leads to a reduction of the series resistance $R_s$, and more importantly it leads to the appearance of a parallel shunt $R_p$ of finite value (~300 ohm) and a significant reduction in the barrier height from 0.52 to 0.33 eV for V$\geq$V$^*$ and also similar values of the parameters for V < V$^*$. On successive transition to lower resistance states there is a gradual reduction in the barrier height. For the final state (LSR$_f$), that is stable for V> V$^*$, the barrier reduces to as low as ~0.22eV. The series resistance $R_s$, also comes down to a low value (7ohm). Interestingly, as can be seen from Table I, the shunt resistance $R_p$ after it forms at the first stage (formation of the LSR$_1$ state) remains constant at $R_p$ ~280-300 ohm. The "switching" back of the successive final low resistance state LRS$_f$ to the initial state LSR$_1$ is seen through the fact that below V=V$^*$ (~0.8V) the I-V curves can be fitted with the value of the barrier height that is close to the barrier height (0.35eV) of the LSR$_1$ state.

In the reverse bias, the stable state is the LRS$_1$ state and the junction current in this state in presence of the shunt is carried by the shunt only because the diode junction carries negligible current in reverse bias. The junction in reverse bias, after it makes transition to the LRS$_1$ state, is dominated by the shunt behaves as an ohmic resistor.

The two principal conclusions that we derive from the model are: that appearance of a shunt resistance $R_p$ that is more or less unchanged after the first switch to the LSR$_1$ state and that there is a lowering of the barrier height on bias recycling that controls the junction current.

### V.2 PHYSICAL PROCESS:

The appearance of a shunt resistance $R_p$ after the first transition to the $LSR_1$ state and which is retained can happen due to appearance of conducting filaments on application of a minimum bias needed for its formation (which in this case is V~1.9V). Appearance of such a filament has been proposed in the case of unipolar switching in binary oxides.[18,19] To check for appearance of such filaments we have taken cross-sectional TEM of the devices. This is shown in Figure 7. It can be clearly seen that when the shunt resistance of finite value appears there are clear bridges (filaments) that connect the two electrodes. These filaments once formed stay within the device and make the $LSR_1$ state a stable state mainly controlled by the shunt.

In the following we propose a likely scenario for the reduction in the barrier and appearance of the shunt resistance. We relate it to the motion of oxygen vacancy /ions on application of the bias. As stated before, electro migration of oxygen is emerging as a very important mechanism for modification of oxide junctions.[21] It has been known for some time that application of even a moderate bias in oxides can make the oxygen (and the oxygen vacancy) mobile in the perovskite oxides.[14] Modification of oxides due to voltage bias has been directly imaged in binary oxide based junctions using $CeO_2$.[16] In the PCMO/Nb:STO junction the hole doped PCMO acts as p-type semiconductor while electron doped Nb:STO acts as n-type semiconductor. Thus application of a positive bias to PCMO makes the junction forward bias. In both of them the carrier density can be quite large (~$10^{18}$-$10^{19}$ /cm$^3$). In Nb:STO can have large O-vacancy ($O_v$) concentration near the surface. Application of a positive voltage to PCMO will attract $O^{2-}$ ions towards it. The bias is applied by the silver contact and there will be migration of $O^{2-}$ ions towards it. This will make the PCMO more conducting by reducing the concentration of oxygen vacancy ($O_v$) in it, which acts like a compensation for the hole doped PCMO. The bias is applied with the Ag top electrode and thus there will be migration of $O^{2-}$ towards the Ag top electrode.

This will have two effects. The $O^{2-}$ migration need not happen uniformly across the film area. The migration involves barrier crossing and will happen along paths where the barriers are low. This can lead to creation of conducting channels which will act as shunt resistance as has been observed in the TEM image shown in Fig. 7. In addition, the movement of Oxygen away from the junction region will create Oxygen vacancy near the Nb:STO surface. Nb:STO being n-type, creation of such oxygen vacancy will create more carriers in the barrier regions and will reduce the barrier height in the diode. The substantial reduction of the barrier height of the junction, as has been observed in the model fit, is a manifestation of the phenomena. The modification of the barrier region occurs locally at the junction region that will involve localized migration while creation of conducting channels that make the shunt will involve migration over a log distance ~500nm.

### V.3 IMPLICATION ON SWITCHING:

The switching devices made from oxides, as stated earlier show two classes of behavior. Generally unipolar switching are seen in binary oxides,[18,19] which form one class of device while the perovskite oxides like PCMO which show bipolar switching behavior[22] form yet another class.

In our case the combination of the diode junction and the shunt appears to produce a situation where at a voltage $V=V^*$, the device reproducibly switches between two stable states $LRS_f$ and $LRS_1$. At a given bias (~2V), the two states can differ in resistances by a factor of 5. While there is a change of resistance state and the device can switch between two states, this device, however, cannot form a memory element as there is no hysteresis and the switching between the two states occurs at the same bias. Nevertheless, the demonstration that one can have an initial forming state and then bias cycling can change the PCMO/Nb:STO junction in a way that it develops transition to lower resistance state and relaxation to the initial formed state with unipolar bias, is an important development towards unipolar switching in complex perovskite oxides.

To conclude, we have shown that a pristine oxide junction PCMO/NB:STO that behaves as a p-n junction, can be formed into a permanent state of lower resistance due to development of shunt across the device thin film (PCMO). On successive voltage cycling this can be transformed into states of successive lower resistance that can be transformed back to the initial stable state below a certain bias. A simple model to understand the junction has shown that this happens due to lowering of barrier at the interface and also reduction in series resistance. It is suggested that this is related to migration of oxygen ions and vacancies at the junction region that enhances majority carrier and thus reduces the barrier height. The initial state that forms after the pristine p-n junction is first taken through a voltage cycle (the step that forms the shunt) appears to occur due to formation of filamentary bridges across the thickness of the PCMO film stretching to the junction region.

**ACKNOWLDGEMENT**

B.G. wants to thank DST (Govt. of India) for the project under Women Scientist Scheme (SR/WOS-A/PS-15/2008).

## FIGURE CAPTIONS :

**FIG. 1**. X-ray diffraction (XRD) data of oriented film of PCMO grown on Nb:STO substrate.

**FIG. 2. (color online)** **(a)** Cross sectional SEM image of the PCMO/Nb:STO device with Ag top electrode. The inset shows the schematic of the device structure and probe placement during measurements Arrow indicates the current direction.**(b)** Ag top electrode of lateral dimension 200μm deposited on PCMO/Nb:STO film by evaporation through metal musk.

**FIG.3.** The in-plane I-V data (using 4-probe arrangements) taken on the PCMO/STO film in the voltage range upto ±4V.

**FIG.4. (color online)** I-V curve (up to V < 1.5V) for the virgin PCMO/Nb:STO devices (PCMO positive) before start of the voltage cycling. The solid line through the data is the fit to the model of a p-n junction with a series resistance.

**FIG.5.(color online)** Effect of voltage cycling with the maximum voltage excursion ($V_{max}$= 2V, 2.5V & 3V). The voltage bias was scanned as $0\rightarrow +Vmax\rightarrow0\rightarrow -Vmax\rightarrow0$ V.

a) Voltage cycling with $V_{max}$=2V, virgin diode like state switches to lower resistance state ($LRS_1$) at a forward bias of V=1.9V and stays at $LRS_1$ on voltage cycling ($V_{max} \leq 2V$) including swing to reverse bias.

b) Gradual enhancement of voltage ($V_{max}$) leads to transition to successive lower resistance states. On reducing the bias the lower resistance states ($LRS_1$, $LRS_2$, $LRS_3$ etc.) changes back to the $LRS_1$ state at V≈$V^*$and for V < $V^*$ including the sweep in the reverse bias region, the I-V curve follows that of the $LRS_1$ state.

**FIG.6.(color online)** I-V curve of the final low resistance state ($LRS_f$) obtained with $V_{max}$ =3V along with that of the I-V curve of the $LRS_1$ state of the device. Solid line indicates the fitting

curve to the model (see inset) of a p-n junction diode with a series resistance ($R_s$) and a shunt resistor ($R_p$).

**FIG.7**. Cross-sectional TEM image of the PCMO/Nb:STO device: Appearance of clear filamentary bridges connecting two electrodes can be seen.

**TABLE CAPTION:**

**Table.1.  Parameters obtained by fitting I-V data taken for different values of $V_{max}$ to the proposed model.**

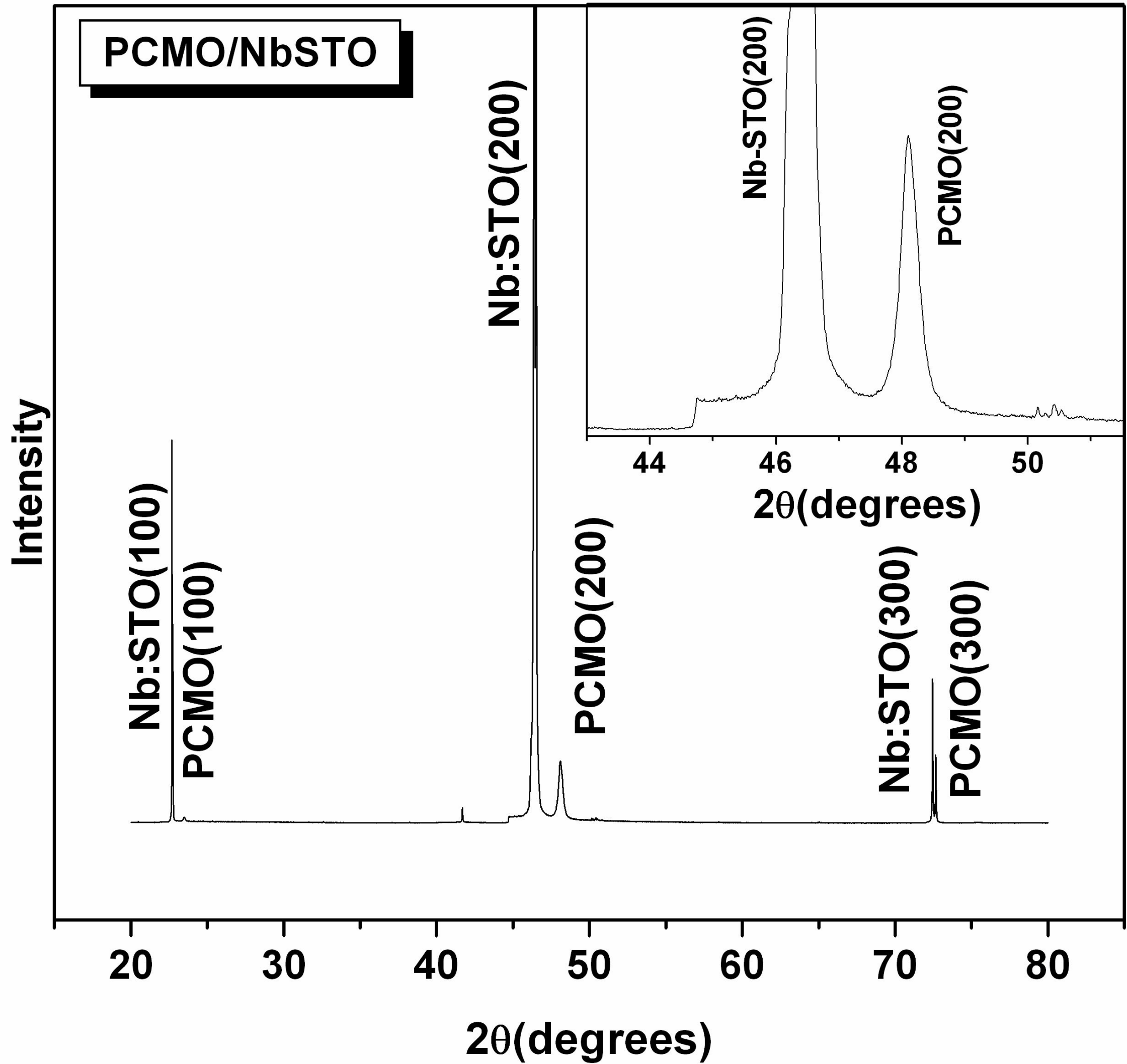

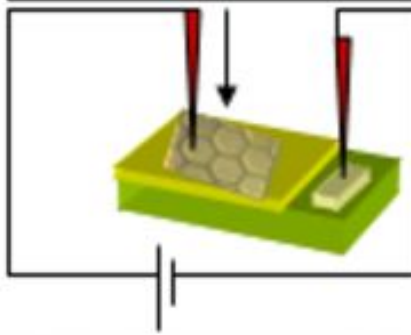

**(a)**

Ag

PCMO

500nm

Nb:STO

Ag/PCMO/Nb-STO

| HV | WD | det | mag ⊞ | 5/28/2010 |
| 25.00 kV | 4.2 mm | TLD | 50 000 x | 5:01:44 PM |

— 1 μm —

**(b)**

| HV | WD | det | mag ⊞ | 4/8/2010 |
| 5.00 kV | 4.0 mm | ETD | 200 x | 5:21:46 PM |

— 200 μm —

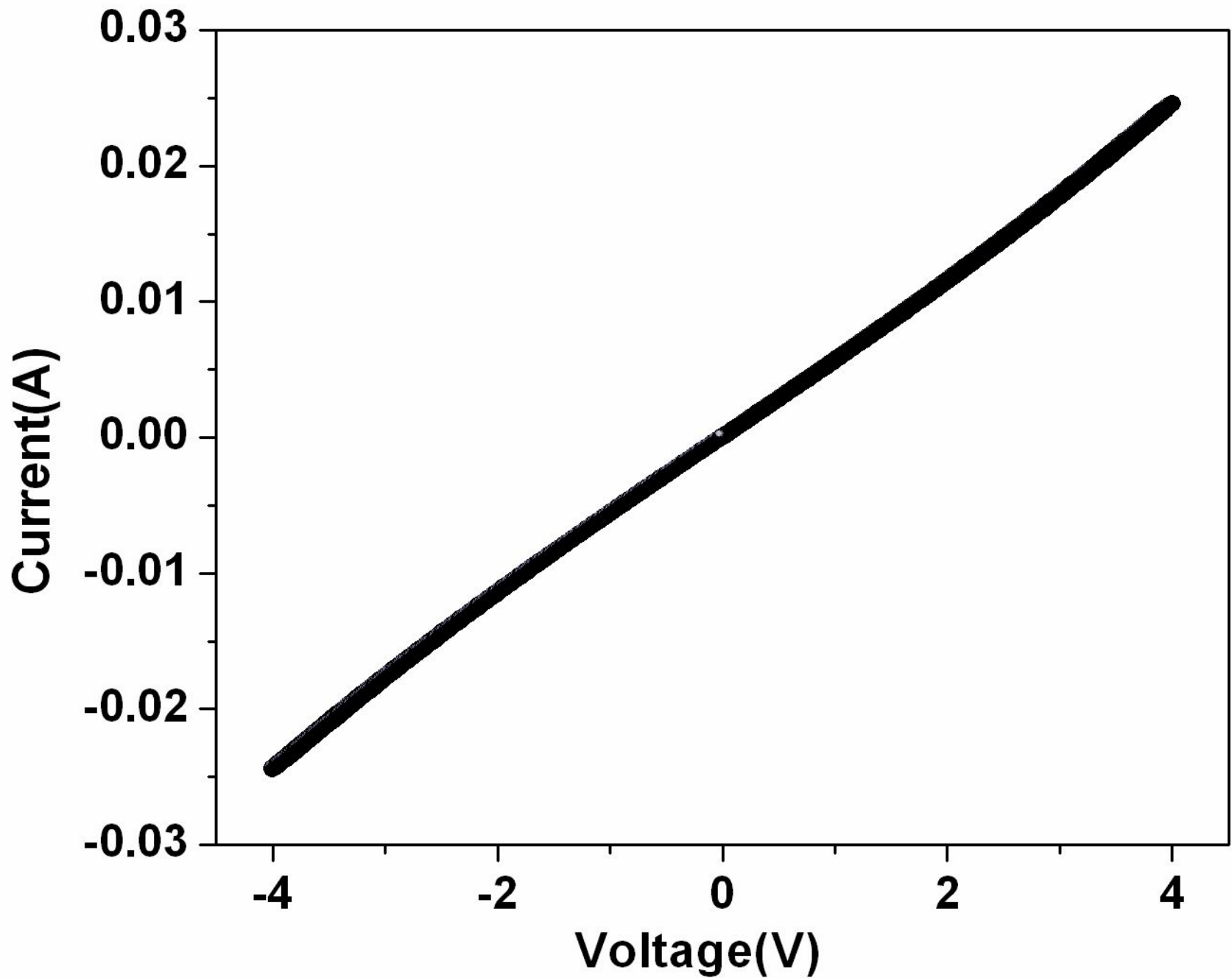

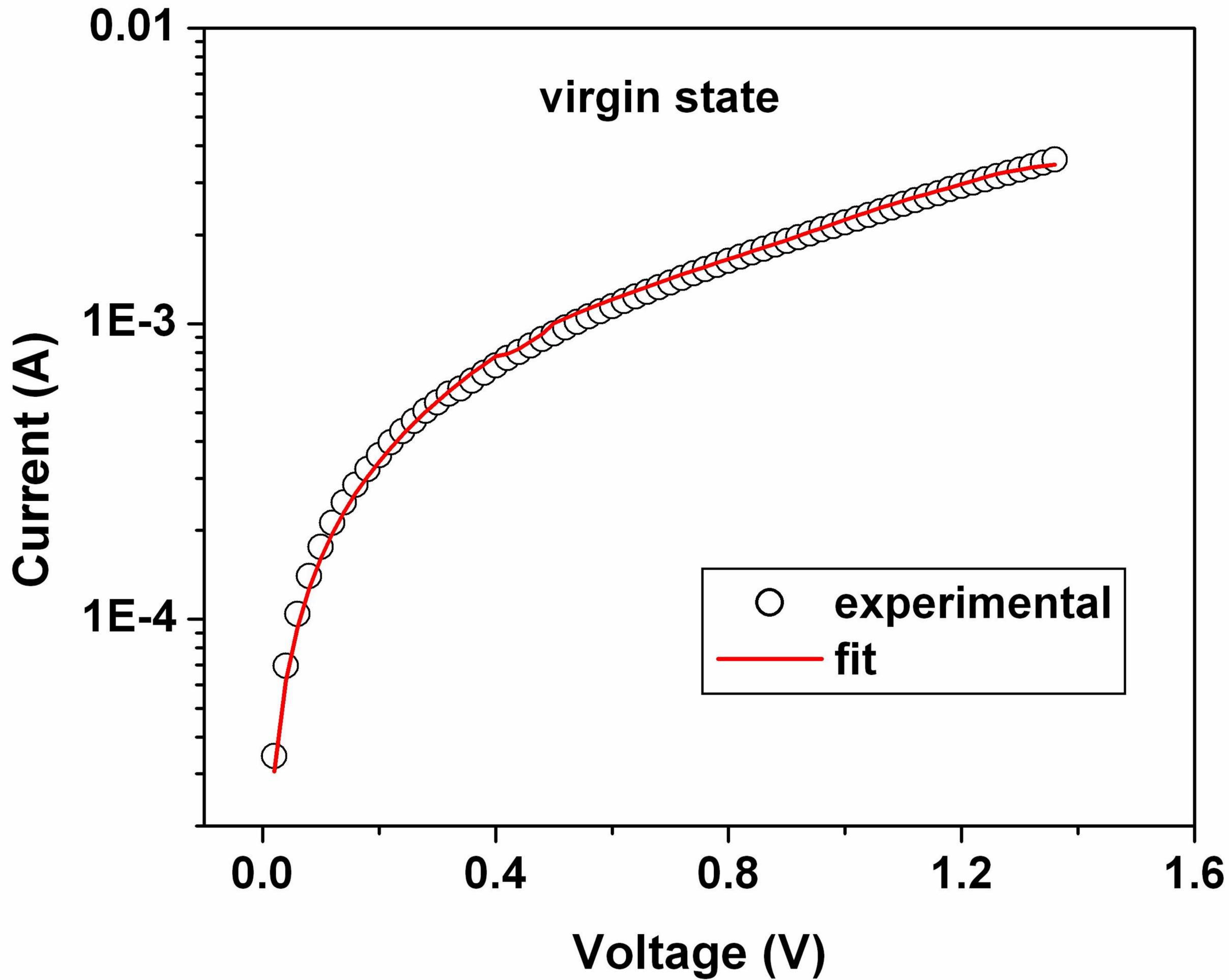

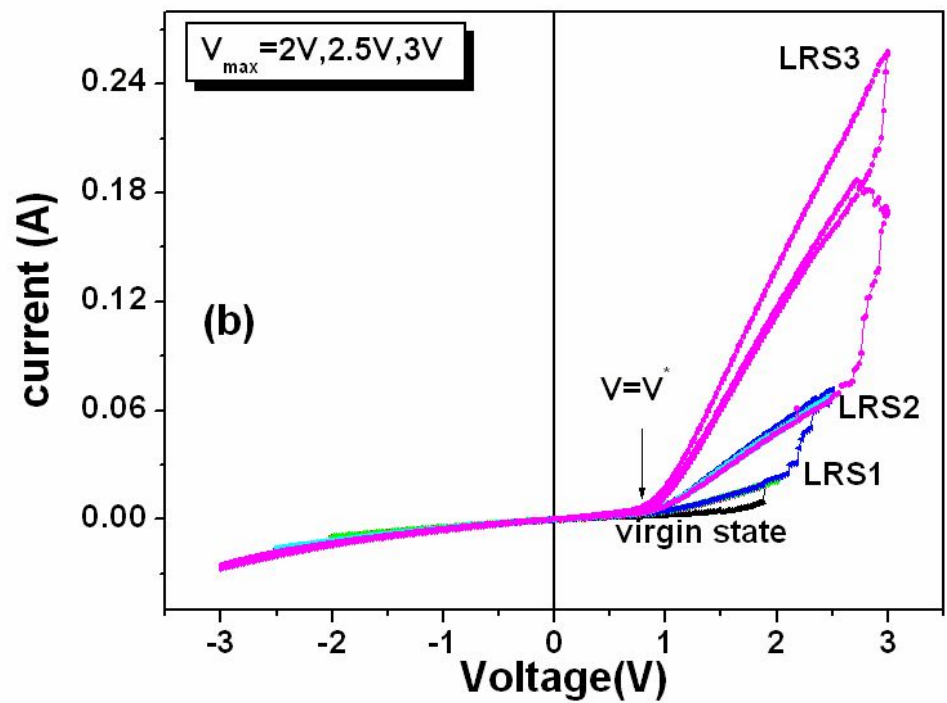

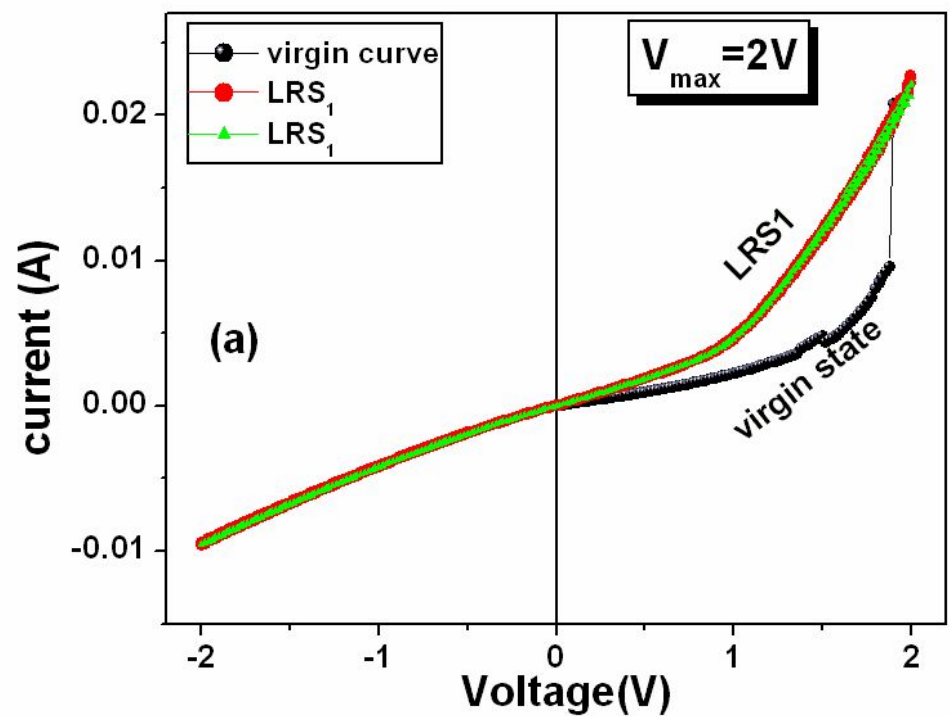

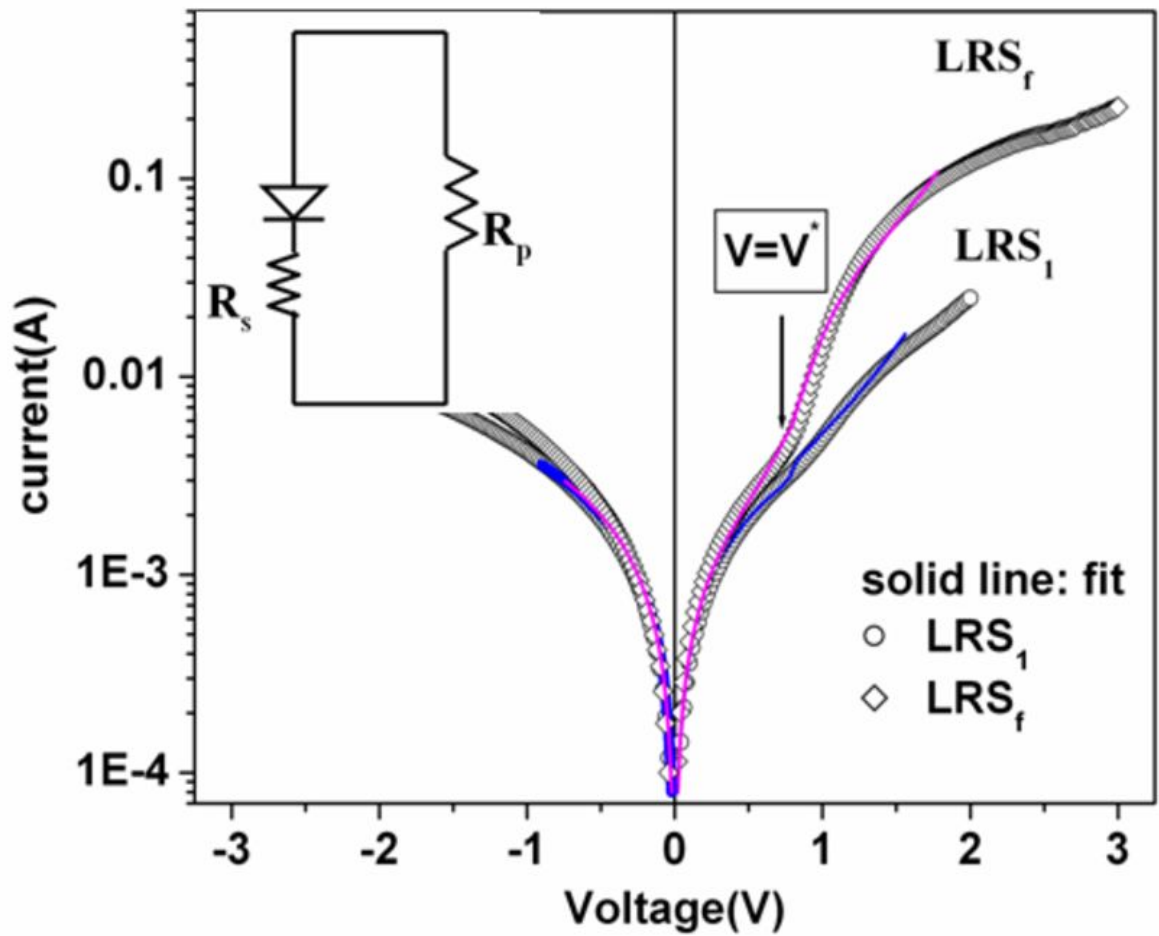

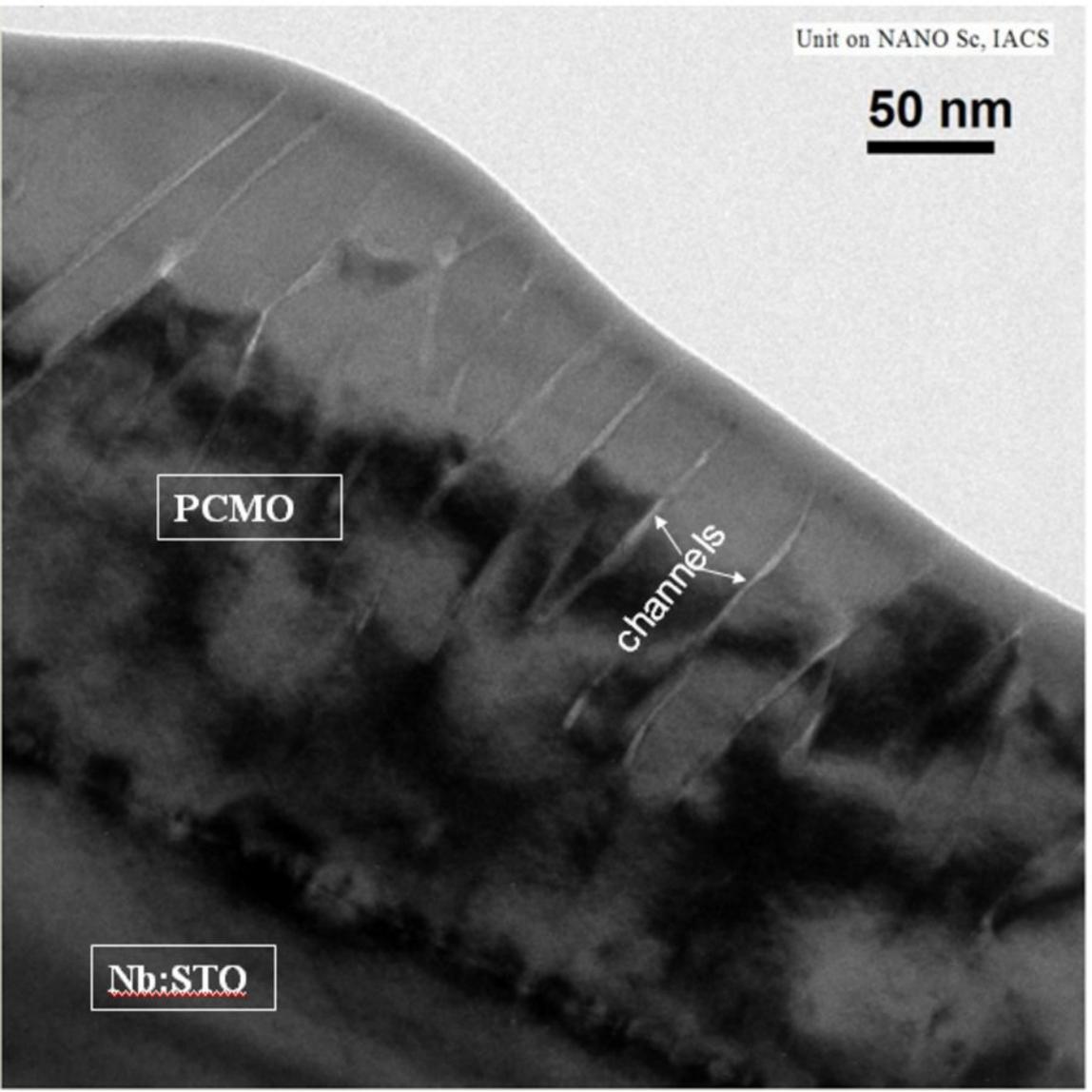